\documentclass[12pt,preprint]{aastex}

\slugcomment{to appear in ApJ}
\shortauthors{Hoard et al.}
\shorttitle{Mid-IR Spectrum of EF Eri}

\begin{document}

\title{The Mid-Infrared Spectrum of the Short Orbital Period Polar 
EF~Eridani from the Spitzer Space Telescope}
 
\author{D. W. Hoard\altaffilmark{1},
        Steve B. Howell\altaffilmark{2},
        Carolyn S. Brinkworth\altaffilmark{1},
        David R. Ciardi\altaffilmark{3},
        Stefanie Wachter\altaffilmark{1}
}
\altaffiltext{1}{Spitzer Science Center, California Institute of Technology,
Pasadena, CA 91125}
\altaffiltext{2}{WIYN Observatory and National Optical Astronomy Observatory,
Tucson, AZ 85719}
\altaffiltext{3}{Michelson Science Center, California Institute of Technology,
Pasadena, CA 91125}

\begin{abstract}
We present the first mid-infrared (5.5--14.5 $\mu$m) spectrum of 
a highly magnetic cataclysmic variable, EF Eridani, obtained with 
the Infrared Spectrograph on the {\em Spitzer Space Telescope}.  
The spectrum displays a relatively flat, featureless continuum.  
A spectral energy distribution model consisting of a 9500 K white 
dwarf, L5 secondary star, cyclotron emission corresponding to a 
$B\approx13$ MG white dwarf magnetic field, and an optically thin 
circumbinary dust disk is in reasonable agreement with the extant 
2MASS, IRAC, and IRS observations of EF Eri.  Cyclotron emission 
is ruled out as a dominant contributor to the infrared flux 
density at wavelengths $\gtrsim3$ $\mu$m.  The spectral energy 
distribution longward of $\sim5$ $\mu$m is dominated by dust 
emission.  Even longer wavelength observations would test the 
model's prediction of a continuing gradual decline in the 
circumbinary disk-dominated region of the spectral energy distribution. 
\end{abstract}


\section{Introduction}

Cataclysmic variables (CVs) are interacting binary stars 
containing a white dwarf (WD) primary and a low mass 
secondary \citep{warner95}.  The evolution of these 
binaries is believed to proceed as follows (see 
\citealt{howell01}):\ after a common envelope phase 
following the post-main sequence evolution of the WD 
progenitor, the low mass secondary star eventually 
(over)fills its Roche lobe and the binary commences 
mass transfer.  Due to angular momentum losses, primarily 
via magnetic braking and gravitational radiation, the 
two component stars move closer together over time; that 
is, their orbital period decreases.  For the oldest CVs, 
the orbital periods are very short (near 80 minutes) and 
the secondary stars are very low mass, being 
$\sim$0.06 M$_{\odot}$ stars or lower mass degenerate 
brown dwarf-like objects.

EF Eridani contains a strongly magnetic WD ($B\approx13$--14 MG; 
\citealt{WR98,howell06b,beuermann07}), making it a member of the 
polar class of CV (named for the highly polarized nature of 
their emitted light).  Unlike CVs containing non-magnetic WDs, 
polars have no accretion disks (instead, accretion proceeds 
directly from the inner Lagrangian point onto the magnetic 
field lines of the WD) and generally do not undergo 
dwarf-nova-type (i.e., disk instability) outbursts. They do, 
however, experience periods of normal mass transfer from the 
secondary star (high states) interspersed with times when 
this accretion flow stops or is significantly decreased 
(low states), possibly related to stellar activity on the 
secondary star.  EF Eri has been in a low state for the past 
10 years \citep{howell06b}.  Interestingly, recent high energy 
observations in the UV \citep{szkody06} and X-rays 
\citep{schwope07} have shown that EF Eri still has a 
$\sim20,000$~K hot spot remaining on the surface of its 
$\sim10,000$~K WD even after a decade of very low mass 
accretion ($\dot{M}\sim10^{-14} M_{\odot}$ yr$^{-1}$). 
This hot region is presumably at or near the active accretion 
pole, which is best modelled as a non-uniform spot covering 
10--20\% of one side of the WD \citep{beuermann07}.

Our initial {\em Spitzer Space Telescope} \citep{werner04} 
photometric observations of a small sample of polars using 
the Infrared Array Camera (IRAC; \citealt{fazio04}) included 
EF Eri as the brightest sample member, and revealed a nearly 
flat mid-IR (3.6--8 $\mu$m) flux density level near 700 $\mu$Jy 
\citep{howell06a,brinkworth07}.  In \citet{howell06a}, we showed that 
the IRAC spectral energy distributions (SEDs) of EF Eri and 
three other polars with similar orbital periods 
had flux density in excess of that produced by the two 
component stars alone.  It was concluded that the best 
candidate for the excess emission was a circumbinary dust disk 
with an inner edge temperature ($T_{\rm in}$) near 800~K.  
In \citet{brinkworth07} (hereafter, B07), we 
used these same data, as well as IRAC observations of two 
additional polars (making a total of seven systems), 
to examine a larger suite of more 
sophisticated SED models.  For EF Eri, we 
found that the most plausible way to produce the observed 
bright 8-$\mu$m flux density, without exceeding the observed 
flux densities at shorter wavelengths, was via a geometrically 
thin, optically thick circumbinary disk (CBD) with 
T$_{in}\approx650$~K; however, we could not completely rule 
out the possibility that at least some of the long wavelength 
flux density in the SED is due to cyclotron emission.
The B07 model results were generally consistent with the need
for a prominent cyclotron emission component to explain the 
near-IR (e.g., 2MASS) portion of the EF Eri SED.

The {\em Spitzer} photometric observations alone could not 
provide any further clarification as to the exact origin of 
the 3.6--8-$\mu$m SED, as they had limited wavelength 
resolution and coverage.  Consequently, those data were not 
able to strongly constrain models for the origin of the 
observed mid-IR flux density of EF Eri (as described in more 
detail in B07).  Therefore, in an effort 
to better understand the true nature and extent of the mid-IR 
emitting source(s) in EF Eri, we obtained a spectrum spanning 
5.5--14.5 $\mu$m using the Infrared Spectrograph 
(IRS; \citealt{houck04}) on {\em Spitzer}.

\section{Observations and Data Processing}
\subsection{Mid-IR Spectrum}

Our spectroscopic observations of EF Eri were obtained using 
the ``Short Low'' module of the IRS, which covers 5.2--8.7 $\mu$m 
in second order (SL2) and 7.4--14.5 $\mu$m in first order (SL1) 
at a resolution of $R\sim60$--120.  
A third order (the SL3 ``bonus'' spectrum) is obtained with each 
SL2 observation; it spans 7.4--8.6 $\mu$m and is primarily used 
to ensure that the flux calibration from SL2 to SL1 is consistent 
(in our case, all three orders were in excellent agreement in the 
overlapping wavelength region, so we did not apply any offsets 
between orders, and used the data from all three orders in 
constructing a final spectrum -- see below).
We obtained four cycles of 240~s 
each for both SL1 and SL2 (i.e., a total of sixteen individual 
exposures after counting the two nod positions obtained in each cycle).  
The corresponding AOR reqkey number is 17052928.

We used the Spitzer Science Center post-BCD software SPICE 
(v1.4.1)\footnote{See 
\url{http://ssc.spitzer.caltech.edu/postbcd/spice.html}.} to extract 
the IRS spectra from the two-dimensonal images.  
The input images consisted of the four S15.3.0 pipeline-combined 
post-BCD images (i.e., the $*$bksub.fits files), each of which is 
constructed from four appropriately background-subtracted, masked, 
and co-added sub-exposures to give two images each for the SL1 and 
SL2+SL3 orders (i.e., one image per order at each nod position).  
We extracted the spectra from these two-dimensional images using 
the optimal extraction algorithm with the standard aperture width. 
This results in two extracted spectra (one for each nod position) 
spanning the three SL orders.

Next, we performed a weighted average of the two (or three) points 
at each wavelength from the two nod spectra to obtain a preliminary 
average spectrum.  We calculated the weighted mean, 
$\langle f_{\rm w} \rangle$, and standard 
deviation of the weighted mean, $\sigma_{\rm wavg}$, 
of the flux density over the full 
wavelength range for the preliminary average spectrum.  Then, we 
rejected any point in an individual nod spectrum that was more 
than $3\sigma_{\rm wavg}$ away from $\langle f_{\rm w} \rangle$ 
when the corresponding point(s) at the same wavelength in the
other nod was not (i.e., an outlier rejection).  
Finally, we recalculated the average spectrum using the 
outlier-rejected nod spectra, via a weighted average when two or 
more data points were available for a given wavelength 
or using the remaining data point when 
only one survived rejection.

\begin{figure}
\epsscale{1.00}
\plotone{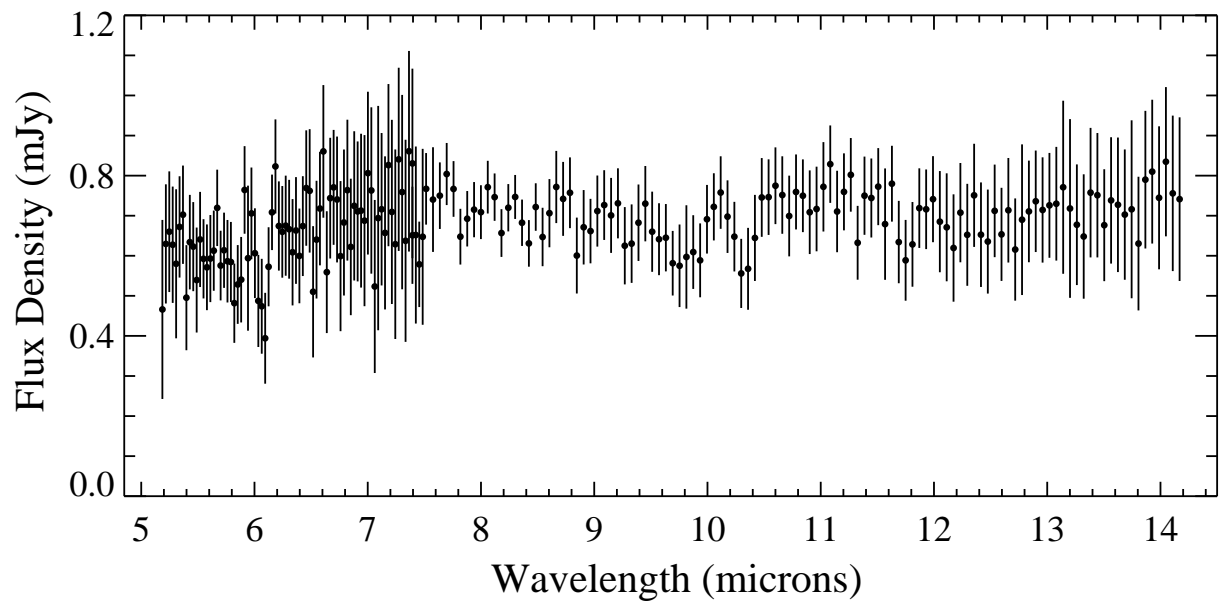}
\epsscale{1.00}
\caption{Mid-IR spectrum of EF Eri from the {\em Spitzer Space 
Telescope} (with $1\sigma$ error bars).
\label{f:spectrum}}
\end{figure}

The IRS spectrum of EF Eri is shown in Figure \ref{f:spectrum}.  
It is characterized by a generally flat continuum with a slight 
dip at the short wavelength end.  There are no obvious emission 
features, and only a few potential absorption features.  However, 
we note that the IRS was designed to achieve high sensitivity 
at the cost of reduced dynamic range; hence, great care must be 
used in the interpretation of weak spectral features\footnote{See 
the IRS chapter of the Spitzer Observer's Manual, available at 
\url{http://ssc.spitzer.caltech.edu/documents/som/}.}, especially 
in spectra of faint targets like EF Eri. 
 The spectrum of EF Eri used in this work has not been scaled 
from its original flux calibration -- the excellent agreement 
in flux density between the IRAC photometric points and the IRS 
spectrum (see \S\ref{s:otherdata} and Figure \ref{f:bigspec}), 
as well as the smooth transition from the SL2 to SL1 data sections 
(at $\lambda\approx7.5$ $\mu$m), attests that the overall continuum 
shape of the spectrum is reliable.  Consequently, our analysis 
presented here will rely on looking at the gross properties of the 
spectrum as a whole (i.e., the continuum shape and flux density 
level), rather than focusing on specific features with low 
statistical probability of being real.

\subsection{Other Data}
\label{s:otherdata}

We also utilized the 2MASS and IRAC photometry of EF Eri reported 
in B07, and the near-IR spectrum shown in Figure 1 of \citet{harrison07}.  
Figure \ref{f:bigspec} shows all of the spectroscopic and photometric 
data plotted together.  Other than the points from IRAC channels 1 
and 2 (3.6 and 4.5 $\mu$m, for which there are no overlapping 
spectroscopic data), all of the photometric data are in 
excellent agreement with the spectroscopic data.  A slight exception 
to this is the 2MASS $K_{\rm s}$-band point, which is significantly 
lower in flux density (by more than $1\sigma$) compared to the 
near-IR spectrum.  This is likely due to the $K$-band variability 
noted by \citet{harrison04}, which has been linked to cyclotron 
emission that varies with orbital phase.

\section{Spectral Energy Distribution Models}
\subsection{The Code}
\label{s:code}

In B07, we introduced our IR SED modeling code for CVs, which we 
applied to photometric data for several polars, including EF Eri.  
We have now made a number of improvements to the modeling code 
for use here.  First, we have modified the code to allow SEDs to 
be calculated at higher wavelength resolution, which is more 
appropriate for comparing to spectral data.  This modification 
has little effect on the WD and CBD components, but provides 
the necessary resolution to resolve cyclotron humps in the 
cyclotron component.  For the models presented here, we have 
used a wavelength increment of 0.05 $\mu$m in the range 1--14.5 $\mu$m.

Second, we have modified the handling of the secondary star to 
accommodate data at longer wavelengths.  In B07, the secondary 
star was represented by average 2MASS and IRAC photometry for 
spectal type templates from \citet{patten06}.  We have now added 
the capability to combine the 2MASS and IRAC photometry of a single, 
representative spectral type star from \citet{patten06} with 
the IRS SL spectrum of the same star from \citet{cushing06}.  
For the EF Eri models presented here, 
the secondary star model component (see Table \ref{t:shared-params}) 
is represented by 
the L5 star 2MASS~J15074769$-$1627386 \citep{reid00} scaled to 
a distance of $d=132$ pc (see \S\ref{s:models}).  

Third, we have made a number of minor improvements to the 
calculation of the optically thin CBD component SED, which is 
primarily used in this work instead of the optically thick CBD used in B07.  
In general, the procedure remains as described in B07.  
The SED of the CBD is obtained by summing the contribution of 
1000 annular rings, each of which has the same width and the 
correct temperature for its average radial distance from the 
center-of-mass of the CV  
(based on the $T \propto [1/r]^{3/4}$ profile explained in 
more detail in B07).
The volume of a ring increases as its radial 
distance increases, and we assume equal mass in each ring, which 
results in rings that are successively less dense at larger 
radial distances.  

We have imposed an 
upper limit of 1000 K for the temperature of the inner edge of 
the CBD.  Not only is it difficult to devise a mechanism that 
would heat the inner edge of the CBD in EF Eri above 1000 K, but 
dust sublimation (hence, destruction of the CBD) likely occurs 
at temperatures above 1000--2000 K.
Based purely on the WD effective temperature, we would expect the 
temperature at the radius of the inner edge of the CBD to be 
$T_{\rm in}\approx250$ K; including the contribution from the 
secondary star, this temperature increases to 
$T_{\rm in}\approx400$--500 K.
However, as noted in B07, the upper limit 
to this temperature is difficult to assess because of the 
uncertainties in how to properly account for the 
contributions from the secondary star and accretion luminosity; 
in addition, the conversion from the ambient temperature of 
irradiation at a particular radius to temperature of the CBD 
material at that radius is uncertain and highly dependent on the 
physical parameters of the CBD and the WD accretion spot(s).
Especially considering the evidence from recent UV and X-ray 
observations \citep{szkody06,schwope07} that there is a 
significant hot spot on the WD in EF Eri, the temperature at 
the inner edge of the CBD could be up to 
several hundreds of K higher.  Effectively, then, we define an 
``allowed'' range of $T_{\rm in}=400$--1000 K. 

Finally, and most significantly, we have replaced the purely 
morphological, template-based cyclotron component from B07 with 
a physical model based on the formulation discussed and used in, 
for example, \citet{chanmugam80,TC87,schwope90} and references therein.
The calculation of the cyclotron SED is influenced by several 
parameters:\ the WD magnetic field strength ($B$), the viewing 
angle ($\theta$), the electron temperature ($kT$), a dimensionless 
``size'' parameter ($\Lambda$), and a normalization factor ($A$).  
None of these parameters is strictly independent from the 
others -- the change in the cyclotron SED produced by adjusting 
one parameter can usually be offset by adjusting one or more of the 
other parameters.  This makes it difficult to produce a unique 
solution without having reliable, very narrow constraints for as 
many of the parameters as possible.  We will describe the constraints 
used to narrow down the possible cyclotron SEDs for EF Eri 
in \S\ref{s:models}.  In the remainder of this section, we will 
discuss, in general terms, the manner in which each parameter 
influences the cyclotron SED.

To first order, the WD magnetic field strength ($B$) determines 
the wavelengths at which cyclotron humps appear in the SED for 
successively higher harmonic numbers.  Smaller values of $B$ 
produce more redshifted humps.  However, the wavelengths of the 
harmonics are also influenced by the electron temperature ($kT$) 
and the viewing angle ($\theta$).  Increasing $kT$ or decreasing 
$\theta$ redshifts the humps.
The humps become broader with increasing $kT$, increasing size 
parameter $\Lambda$ (however, see below for additional effects 
of changing $\Lambda$), or decreasing $\theta$.  At very high 
values of these parameters (especially $kT$), the humps are so 
broad as to be completely overlapping and blended together, 
effectively forming a humpless pseudo-continuum.  At very low 
values, the cyclotron humps are very narrow, resulting in the 
cyclotron spectrum consisting of multiple discrete hump profiles 
with regions of zero intensity between them.
The normalization factor ($A$) is a simple scaling factor given 
by the ratio of the effective emitting area of the cyclotron 
component to the square of the distance to the CV.

The size parameter $\Lambda$ deserves additional explanation.  
Although its value is set in the cyclotron SED calculation as 
a single number, this parameter actually incorporates several 
other parameters.  It is defined (see \citealt{schwope90}) as
\begin{equation}
\Lambda = 2.01\times10^{5} \left(\frac{s}{10^5\,{\rm [cm]}}\right) \left(\frac{N_{\rm e}}{10^{16}\,{\rm [cm^{-3}]}}\right) \left(\frac{3\times10^7\,{\rm [G]}}{B}\right),
\end{equation}
\noindent where $s$ is the geometric path length through the 
cyclotron emitting region and $N_{\rm e}$ is the electron number 
density.  For a fixed value of $B$, the value of $\Lambda$ is 
dependent on only the product of $s$ and $N_{\rm e}$.  
Functionally, larger values of $\Lambda$ correspond to higher 
optical depths, which influences the harmonic number at which 
the cyclotron humps transition from being optically thick 
(at lower harmonics) to optically thin (at higher harmonics).  
Optically thick cyclotron humps have a truncated, flat-topped 
appearance compared to the rounded optically thin humps.  
The value of $\Lambda$ also has a large effect on the relative 
amplitudes of the cyclotron humps.  The lower harmonic, optically 
thick humps at longer wavelengths rapidly decline in amplitude as 
harmonic number decreases compared to the optically thin humps at 
higher harmonics (shorter wavelengths).

\subsection{The Models}
\label{s:models}

\subsubsection{Distance to EF Eri}

The distance used to scale the model components in this work is 
the $1\sigma$ upper limit of a trigonometric parallax-derived 
distance from \citet{thor03}, and is approximately halfway between 
the two possible nominal parallax distance values of 113 and 162 pc 
reported in that work.  We increased the distance from the 105 pc 
used in B07 (which was based on non-parallax estimates) because 
it is now clear that cyclotron emission makes a non-negligible 
contribution in the $J$ band (see discussion of models below), 
whereas at the smaller distance, the observed $J$-band flux 
density was accounted for completely by the WD and secondary star 
components.  The value of 132 pc used here is the minimum distance 
for which the summed WD, secondary star, and cyclotron emission 
do not exceed the observed photometric and spectroscopic data 
at $J$ band.

\subsubsection{Cyclotron Component Constraints}
\label{s:cyccon}

As noted in \S\ref{s:code}, our revised cyclotron model component 
is calculated using a number of non-independent parameters, so it 
is helpful to constrain, as much as possible, the range of valid 
parameter space.  To that end, we describe in this section the 
various constraints that we used to limit the cyclotron model 
parameters. 

\begin{figure}
\epsscale{1.00}
\plotone{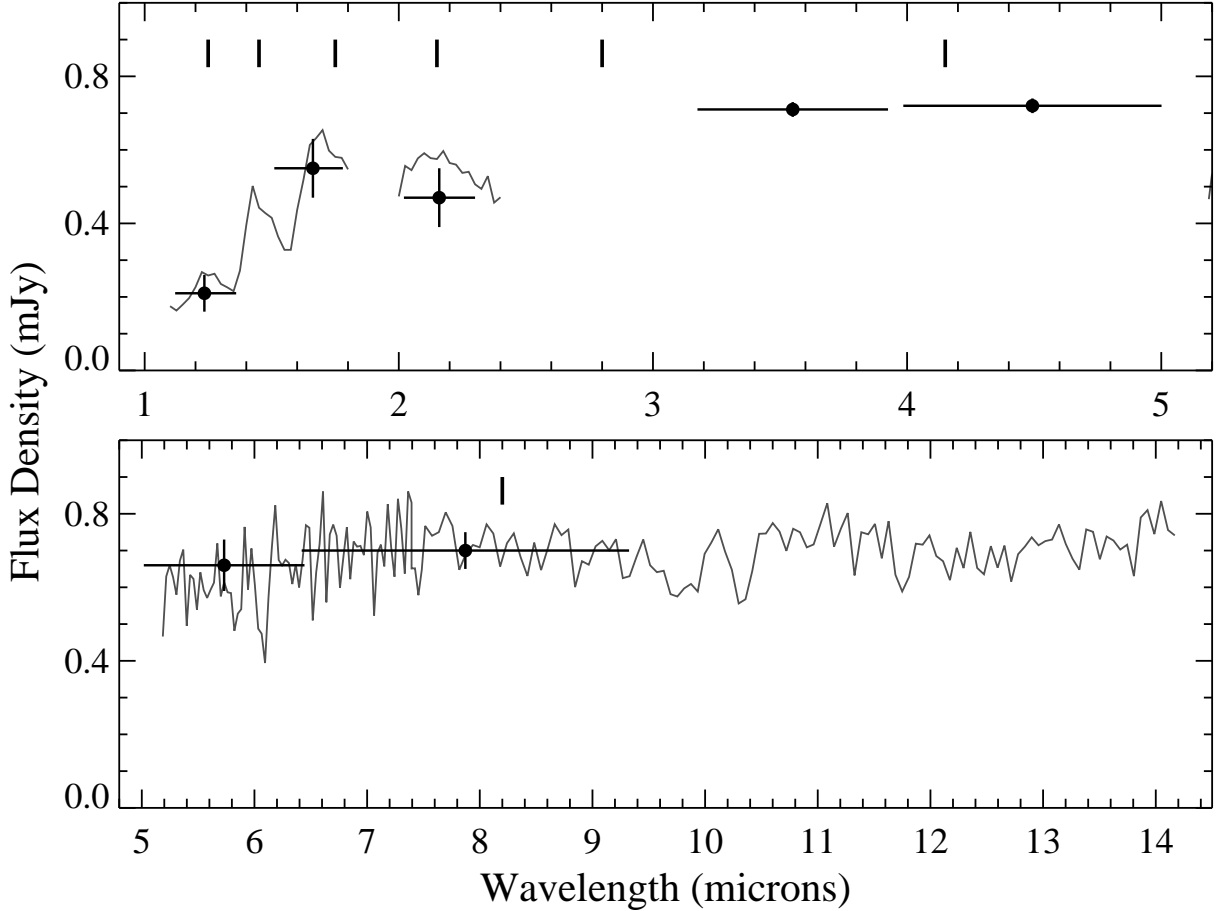}
\epsscale{1.00}
\caption{IR observations of EF Eri.  The photometric data from 
\citet{brinkworth07} are shown as filled circles; from short to 
long wavelength, these are 2MASS $J$, $H$, and $K_{\rm s}$, and 
IRAC channels 1--4.  Error bars on all of the photometric points are 
the $1\sigma$ photometric uncertainties in the y-direction (these 
are smaller than the plotting symbols for IRAC channels 1 and 2), 
and the widths of the photometric bands in the x-direction.
The near-IR spectrum (taken from Figure 1 of \citealt{harrison07}) 
and IRS mid-IR spectrum of EF Eri are shown as solid lines (error 
bars have been omitted for clarity).  
The vertical hashmarks show the cyclotron harmonic peaks for $n=2$--7 
(top panel, right to left) and the cyclotron fundamental (bottom panel)
for the model shown in Figure \ref{f:models}b and
discussed in \S\ref{s:newmodels}.
Note the different wavelength scales in the two panels.
\label{f:bigspec}}
\end{figure}

\begin{description}
\item[$B$:] The WD magnetic field in EF Eri is frequently derived 
as $B=13$--$14$ MG (e.g., \citealt{WR98,howell06b,beuermann07}).  
A value of $B$ in this range is consistent with the location of 
the cyclotron humps in the near-IR spectrum of EF Eri 
(\citealt{harrison07}; also see Figure \ref{f:bigspec}).  
We do not consider the much more complex scenario in which the 
cyclotron spectrum of EF Eri results from two or more magnetic 
accretion regions with significantly different field strengths.

\item[$kT$:] The near-IR spectrum of EF Eri also provides useful 
constraints for this parameter.  Even after subtracting the WD 
and secondary star components, there is residual flux between the 
cyclotron humps.  This requires $kT\gtrsim5$ keV (at lower values 
of $kT$ and for reasonable values of $\Lambda$ -- see below -- the 
cyclotron spectrum is composed of discrete humps with zero flux 
level between them).  
In addition, for $kT\lesssim5$ keV, and for values of $\Lambda$ 
that reproduce the observed optically thick to thin transition 
and relative hump peak amplitudes, the cyclotron humps are too 
narrow compared to the observed near-IR spectrum.
The fact that discrete cyclotron humps are observed requires 
$kT\lesssim20$ keV.  

\item[$\theta$:] In the absence of consistent information 
constraining this parameter, we chose to set $\theta=75^{\circ}$ 
for all models.  In any case, $\theta$ cannot be much lower or 
higher than this, or the widths and relative amplitudes of the 
optically thin humps will not match the observed near-IR spectrum.

\item[$\Lambda$:] For any given values of the other parameters, 
$\Lambda$ is the most constrained parameter.  This is because 
only a narrow range of its values will reproduce the observed 
near-IR spectrum, which shows the $n=4$ harmonic as optically 
thick, and higher harmonics as optically thin.  The value of 
$\Lambda$ is further fine-tuned by matching the relative peak 
amplitudes of both the optically thick and thin cyclotron humps.

\item[$A$:] The main constraint on the scaling parameter is that 
for a given distance, $A$ should correspond to an effective 
emitting area of the cyclotron radiation ($a_{\rm cyc}$) that 
is (much) smaller than the projected surface area of the WD.
\end{description}

Most of these constraints are derived from the near-IR spectrum 
of EF Eri, which shows the cyclotron hump at the $n=4$ harmonic 
to be flat-topped and much lower in peak amplitude than the humps 
at the $n\geq5$ harmonics (see Figure \ref{f:bigspec}).  
This indicates that the transition 
from optically thick to thin cyclotron emission occurs in the 
vicinity of the $n=4$ harmonic.  In particular, satisfying the 
observed transition from optically thick to thin humps results 
in the mid-IR contribution of the cyclotron emission being 
increasingly negligible at longer wavelengths.  That is, the 
cyclotron component is only important at relatively short IR 
wavelengths, and does not contribute significantly to the mid-IR 
region spanned by the Spitzer data.

As expected from the description in \S\ref{s:code}, we found it 
difficult to determine a unique, ``best'' solution for the 
cyclotron model component.  We determined the relative goodness 
of different model cyclotron components by calculating the 
$\chi^2$ and standard deviation of the residuals ($\sigma_{\rm res}$) 
in the 1--2.5 $\mu$m wavelength region for the summed cyclotron, 
WD, and secondary star model components compared to the observed 
near-IR spectrum.  
Table \ref{t:cycmodels} lists several ``best'' cyclotron models 
determined in this way for a range of $kT$ values.  
As $kT$ decreases, we had to decrease $B$ and increase $\Lambda$ 
in order for the cyclotron component to continue to match the 
wavelength spacing, widths, and relative peak amplitudes of the 
observed cyclotron humps.
The goodness of the model also tends to decrease with decreasing 
$kT$, although none of the models has particularly poor agreement 
with the observed SED.
In the end, for the purposes of this work, the specific cyclotron 
component that we use is not particularly important since all of 
them contribute negligibly at mid-IR wavelengths.  We will use 
the $kT=10$ keV cyclotron model in the remainder of this work.

In B07, we utilized a ``sum-over-fields'' approach to calculating 
the cyclotron component, which considered the summed contributions 
from cyclotron emission of electrons encountering a successively 
stronger magnetic field as they approached the WD.  In part, this 
was an attempt to set a ``worst case'' limit for the contribution 
of cyclotron emission at long wavelengths (since the strength of 
cyclotron emission at long wavelengths is increased relative to 
short wavelengths through this approach).  However, another effect 
of the sum-over-fields approach is to smear out the individual 
cyclotron humps, which is clearly inconsistent with the observed 
near-IR spectrum of EF Eri.  Consequently, we have not used that 
approach here.  In any case, even if we considered a two-part 
cyclotron component consisting of the single-field {\em and} 
summed-fields cases, the contribution of the latter would have 
to be extremely small in order to not dilute the strong observed 
single-field spectrum (and, regardless of strength, would not 
contribute significantly at $\lambda\gtrsim4$ $\mu$m).

\subsubsection{Comparison with Brinkworth et al.\ (2007)}
\label{s:comparison}

Figure \ref{f:models}a shows the observational data from 
Figure \ref{f:bigspec} with a model containing an optically thick 
CBD similar to the best optically thick CBD component from B07 
(see Model 1 in Table 4 of that work).  It has been adusted 
slightly to account for the larger distance used here by changing 
the inner edge temperature from 655 K to 755 K.  
The criterion used in B07, of best reproducing the 8-$\mu$m IRAC 
point without exceeding any shorter wavelength point, has been 
preserved.  This model has the wrong spectral shape and 
significantly exceeds the {\em longer} wavelength SED of EF Eri 
that is revealed by our IRS spectrum.

\subsubsection{New Model Results}
\label{s:newmodels}

\begin{figure}
\epsscale{1.00}
\plotone{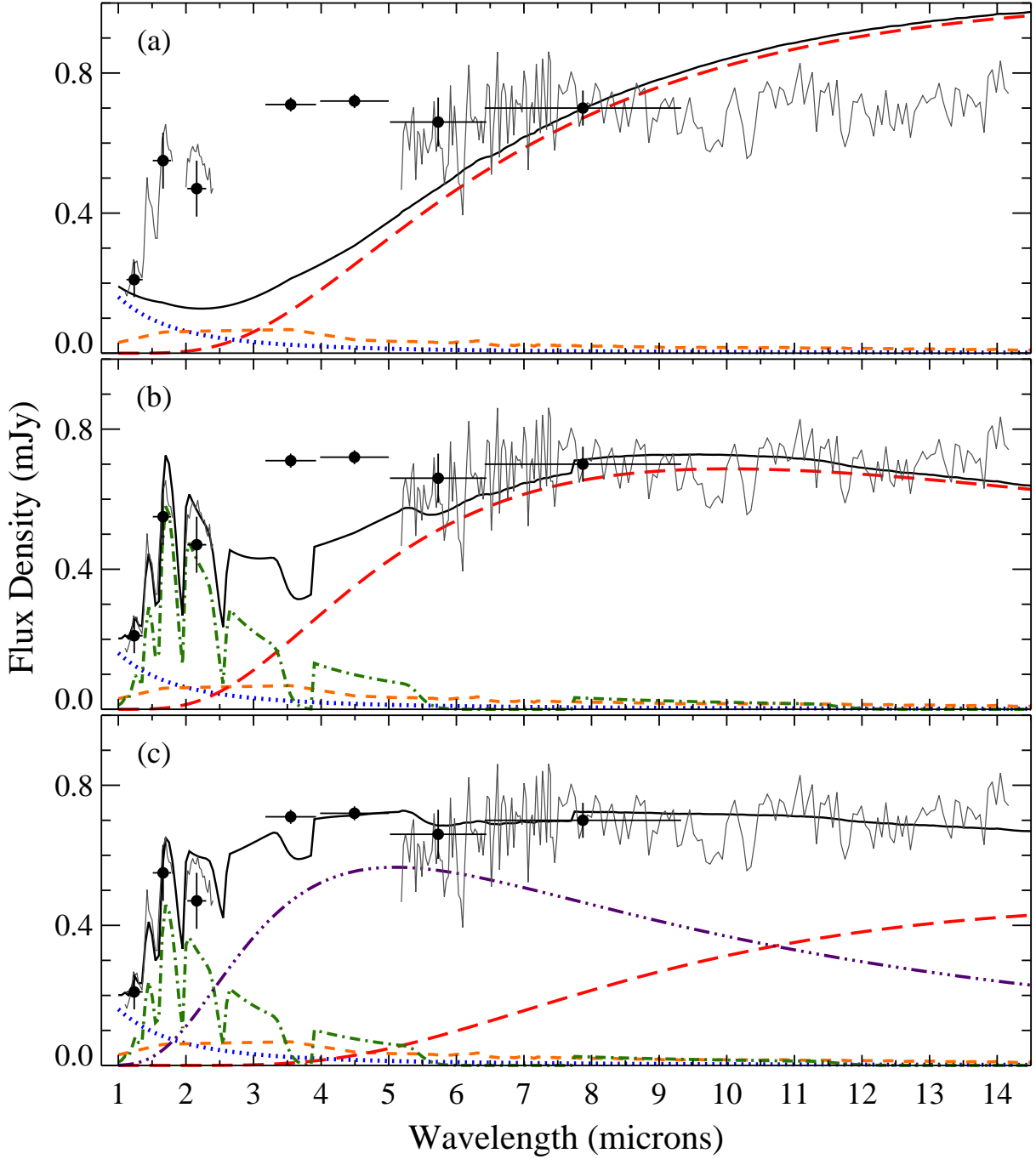}
\epsscale{1.00}
\caption{Mid-IR SED of EF Eri (from Figure \ref{f:bigspec}) with 
models: (a) optically thick CBD model, (b) new optically thin CBD 
and cyclotron model, and (c) new optically thin CBD and cyclotron 
model with an additional blackbody component.  
The model components are: WD (blue dotted line), secondary star 
(orange short dashed line), circumbinary disk (red long dashed 
line), cyclotron emission (green dot-dash line), and blackbody 
(purple dot-dot-dot-dash line); the total combined model is shown 
as a thick solid line (black).
See text for additional details of the models.  
\label{f:models}}
\end{figure}

Figure \ref{f:models}b shows a new model that utilizes the 
additional constraints on flux density at long wavelengths 
provided by our IRS spectrum of EF Eri.  
The parameters common to this model and the one discussed 
below are listed in Table \ref{t:shared-params}, while parameters 
specific to this model are listed in Table \ref{t:specific-params} 
(Model 1).  For this model, we have utilized an optically thin CBD 
composed of spherical dust grains that radiate as blackbodies 
according to the radial temperature profile calculated as for 
the optically thick CBD case (see \S\ref{s:code} and B07).  

This model shows significant improvement over that shown in 
Figure \ref{f:models}a, especially at the short and long wavelength 
ends.  However, the model flux density is too low at the middle 
wavelengths (i.e., IRAC channels 1 and 2 at 3.6 and 4.5 $\mu$m).
The total mass of the CBD is $\approx 10^{21}$ g, consistent with 
the finding from B07 that the masses of CBDs in magnetic CVs are 
many orders of magnitude smaller than predicted to be required to 
influence the angular momentum loss history of these systems \citep{taam03}.  
For lower and higher inner edge CBD temperatures, the overall 
flux density level of the observed SED can be matched by increasing 
or decreasing, respectively, the total disk mass (i.e., the number 
of radiating dust grains).  
If the temperature of the inner edge of the CBD is decreased, then 
the match between the model and observed SEDs becomes worse -- the 
CBD profile does not reach peak flux density until an even longer 
wavelength, which exacerbates the problem of missing flux density 
at the IRAC wavelengths.
A higher temperature for the inner edge of the CBD produces a 
better match at the middle wavelengths.  However, the model is 
then too faint at the long wavelength end, since the CBD SED peaks 
and begins to decline at a shorter wavelength.  
(See \S\ref{s:cbdcon} for more discussion of the constraints on CBD 
model parameters.)

We have explored a possible means of reconciling 
this CBD model with the 3.6 and 4.5 $\mu$m data.
The model in Figure \ref{f:models}c is similar to that shown in 
Figure \ref{f:models}b, but uses a blackbody component to account 
for the ``missing'' flux at 3.6 and 4.5 $\mu$m.  The model 
parameters are listed in Tables \ref{t:shared-params} and 
\ref{t:specific-params} (Model 2).  This model has the advantage 
that it requires a CBD with a low inner edge temperature that 
could easily be produced by irradiation from the stellar 
components in EF Eri.
On the other hand, it has the disadvantage that the physical 
origin of the additional component is unclear.  The required 
equivalent emitting area (corresponding to a radius of $46R_{\rm wd}$) 
is too large to be contained in the stellar Roche lobes, which 
points to the CBD.  The required temperature is 1000 K, which is 
somewhat uncomfortably warm from considerations of both the 
origin of the heating and potential destruction of the dust grains.  
However, it might suggest that a more complex radial temperature 
profile in the CBD could produce an SED shape that is more 
consistent with the observations.  We have not explored this 
possibility in detail because it would introduce yet more free 
parameters into our already barely constrained model.

\subsubsection{Uniqueness, Plausibility, and Constraints of the Circumbinary Disk Models}
\label{s:cbdcon}

Much like the cyclotron emission component (see \S\ref{s:code} 
and \S\ref{s:cyccon}), the model CBD SEDs shown in this work are not, 
strictly speaking, unique solutions, in the sense that very similar 
results can be achieved from somewhat different combinations of input 
parameters.  We have tried to minimize this as much as possible by 
constraining plausible parameter ranges based on whatever other 
information, observational data, and reasonable assumptions are available.  
In this section, we describe in detail the justification for the 
constraints that have been assumed in fixing the exponent in the 
radial temperature profile calculation (see \S\ref{s:code}) and 
the inner radius, $R_{\rm in}$, of the model CBD.  We also expand 
upon the discussion of the failure of the optically thick CBD 
model first mentioned in \S\ref{s:comparison}.

In general terms, the influence of the exponent in the radial 
temperature profile can be described as follows:\   
a larger (smaller) exponent in the CBD radial temperature profile 
leads to a steeper (shallower) temperature gradient near the 
inner edge of the disk and overall lower (higher) temperature 
throughout the disk, which corresponds to large (small) dust grains 
that do (do not) cool efficiently.  In practice, we have found that 
an exponent of 3/4, as used here and in B07, produces the most 
viable results in comparison with our observations of CVs.  
Smaller exponents produce CBD SEDs that rise too steeply and are, 
overall, too bright to reproduce the observed SEDs without 
arbitrarily increasing the distances to the CVs to many 
hundreds or thousands of pc.
Larger exponents produce CBD SEDs that rise too shallowly at longer 
wavelengths and are, overall, too faint to match the observed mid-IR 
flux densities without arbitrarily increasing the temperature of 
the disk's inner edge to unrealistic levels.  
For example, we can obtain a model optically thin CBD SED that 
is essentially indistinguishable from that in Model 1 
(Figure \ref{f:models}b) by using a larger radial 
temperature profile exponent (1 instead of 3/4; see \S\ref{s:code}) 
and smaller total disk mass ($4.41\times10^{20}$ g instead 
of $9.55\times10^{20}$ g); however, one objection to this 
approach is that departure from an exponent of 3/4 implies 
dust grains that are not blackbodies \citep{FKR}, whereas this is 
an implicit assumption of the CBD model calculations.

The inner edge radius of the CBD models calculated in this work 
(for both optically thick and thin cases) is fixed 
at $R_{\rm in}=73R_{\rm wd}$, 
which is a lower limit set by the tidal truncation radius of EF Eri.  
Increasing the inner radius of the CBD from this 
value worsens the agreement between the CBD models and the 
observed SED, because the irradiation-induced temperature of 
the inner edge will then be lower.  This effectively removes flux 
from the short IR wavelengths of the model SED, whereas the main 
problem we have in reproducing the observations is that the models 
already have too little flux at short IR wavelengths.  (The outer 
radius of the CBD is calculated to correspond to a temperature of 
20 K according to the radial temperature profile in use, but at 
wavelengths shortward of 15 $\mu$m the resultant profile is 
insensitive to increasing the temperature of the outer radius 
cut-off by as much as an order of magnitude.)
Even if we arbitrarily (and unphysically) decrease the inner 
edge radius of the CBD to $50R_{\rm wd}$, which is barely larger 
than the distance from the CV's center-of-mass to the back 
of the secondary star's Roche lobe ($49R_{\rm wd}$), 
the resultant model CBD SED (for both 
optically thick and thin cases) has only a few percent improvement 
(increase) in flux density at IRAC channels 1 and 2.  Similarly, 
the agreement with the IRAC channels 3 and 4 and IRS data is not 
significantly better than that achieved by the CBD component used 
in Model 1 (Figure \ref{f:models}b).

The optically thick CBD models are parameterized solely by the 
radial temperature profile (including the value of $T_{\rm in}$) 
and size (i.e., inner and outer radii).  As such, the variety 
of possible model optically thick CBD SEDs is rather more limited 
than for the optically thin case.  Changing either the radial 
temperature profile exponent or $R_{\rm in}$ does not yield an 
optically thick CBD model that reproduces the observed SED any 
better than the model shown in Figure \ref{f:models}a 
and discussed in \S\ref{s:comparison}.
For example, using an exponent of 1, we can produce an optically 
thick CBD SED whose shape (i.e., relative intensity at each 
wavelength) is almost indistinguishable from that of the optically 
thin model CBD SED in Model 1.  However, it is overall 15--30\% 
fainter than the optically thin model, so produces much worse 
agreement with the observations.  Smaller exponents produce model 
CBD SEDs that are far too bright, especially at longer wavelengths.
Increasing $T_{\rm in}$ for the optically thick CBD produces more 
flux at shorter wavelengths, but also increases the steepness of 
the SED and the excess flux at longer wavelengths.  Decreasing 
$T_{\rm in}$ begins to flatten the SED, but also makes it overall 
too faint at all wavelengths, especially the short IR wavelengths.
Increasing $R_{\rm in}$ also fails, since (as described above) 
this results in even less flux at the short IR wavelengths.  
In light of the flat shape of the EF Eri SED at wavelengths 
longer than the IRAC regime, we conclude from the behavior of
the optically thick CBD models described in this section that they 
are much less likely than the more flexible optically thin CBD models
as viable explanations of the mid-IR SED of EF Eri.

\section{Conclusions}

Our newly obtained mid-IR spectrum of EF Eri has allowed us to 
further constrain and refine the SED model for this system first 
presented in B07.  Based on the B07 model, which was constrained 
by only 2MASS and IRAC photometric data, we would have expected 
the 8--14 $\mu$m SED of EF Eri to either rise 
(if dominated by an optically thick CBD) or fall (if dominated 
by short wavelength cyclotron emission).  However, the observed 
spectrum defied both of our expectations, by remaining almost 
flat compared to the IRAC data.  This does allow us to eliminate 
cyclotron emission as a dominant component in the mid-IR SED of 
EF Eri beyond $\lambda\approx3$ $\mu$m.  At the same time, we 
also show that a geometrically thin, but optically thick, 
CBD is unlikely as a viable explanation for the spatial 
distribution of dust in EF Eri.  Instead, the 
dust is more likely present as a geometrically and optically thin CBD.  
In all cases, however, there are inconsistencies between our CBD 
models and the observed SED of EF Eri at the IRAC channel 1 and 2 
wavelengths (3.6 and 4.5 $\mu$m) that imply a more complex situation 
is present than represented by our simple CBD models.
As also found in B07, the total mass of dust in the CBD is still 
several orders of magnitude too small to strongly affect the 
secular evolution of CVs in the context of current models of 
angular momentum loss mechanisms in CVs. 

Based on our study of EF Eri, we can make several generalizations 
regarding the mid-IR observational properties of CVs that contain dust.  
First, EF Eri is a low-field polar.  In polars with moderate to 
strong WD magnetic fields of several tens of MG or more, cyclotron 
emission will be shifted to even shorter wavelengths and be even 
less important for understanding the mid-IR SED.
Second, EF Eri contains a very low mass, faint secondary star.  
Longer orbital period systems will have correspondingly more 
massive, brighter secondary stars.  However, in the IRAC bands, 
an M5 dwarf is only $\approx10$ times brighter than an L5 brown 
dwarf \citep{patten06}.  Although the observed 3.6 $\mu$m flux 
density in EF Eri is comparable to that of an M5 star, at 8 $\mu$m 
the observed flux density is $\approx4$ times that of an M5 star.  
So, even in systems containing a more massive secondary star, dust 
emission at a comparable level to that in EF Eri produces a mid-IR 
SED far in excess of the stellar components.
Finally, EF Eri lacks an accretion disk.  In non-magnetic CVs, the 
hot accretion disk will appear in the IR as a Rayleigh-Jeans-like 
tail similar in shape to the WD SED, but likely significantly 
brighter than both stellar components.  However, the maximum 
possible emitting area for the accretion disk is limited by the 
size of the WD Roche lobe.  The CBD, on the other hand, can have 
an emitting area many orders of magnitude larger.  Hence, even in 
the presence of an accretion disk, the system SED at the longest 
mid-IR wavelengths could still be dominated by dust emission.

An extrapolation of our current CBD model for EF Eri to even 
longer wavelengths predicts a continued gradual decline in the 
overall flux density, with (for example) the total flux density 
of EF Eri at 24 $\mu$m being about 90\% of the 8-$\mu$m value.  
Longer wavelength observations (e.g., at the {\em Spitzer} 
Peak-up Imaging 22-$\mu$m or MIPS 24-$\mu$m bands) could test 
this prediction.

\acknowledgments

This work is based in part on observations made with the 
{\em Spitzer Space Telescope}, which is operated by the Jet 
Propulsion Laboratory, California Institute of Technology, 
under a contract with the National Aeronautics and Space 
Administration (NASA).
Support for this work was provided by NASA.
We thank the Spitzer Science Center (SSC) Director for his 
generous allocation of observing time for the NASA/NOAO/{\em Spitzer 
Space Telescope} Observing Program for Students and Teachers. 
The National Optical Astronomy Observatory (NOAO), which is 
operated by the Association of Universities for Research 
in Astronomy (AURA), Inc., under cooperative agreement with the 
National Science Foundation (NSF), has provided many in kind 
contributions for which SBH is grateful. 
This work makes use of data products from the 
Two Micron All Sky Survey, which is a joint project of the 
University of Massachusetts and the Infrared Processing and 
Analysis Center/Caltech, funded by NASA and the NSF. 
CSB acknowledges support from the SSC Enhanced Science Fund 
and NASA's Michelson Science Center.
DWH thanks Axel Schwope for helpful advice on calculating 
cyclotron spectra.


\clearpage

\begin{deluxetable}{llllllll}
\tablewidth{0pt}
\tabletypesize{\footnotesize}
\tablecaption{Representative cyclotron models \label{t:cycmodels}} 
\tablehead{
\colhead{$kT$} & 
\colhead{$B$} & 
\colhead{$\theta$} & 
\colhead{$\log(\Lambda)$} & 
\colhead{$\log(A)$} & 
\colhead{$a_{\rm cyc}$} & 
\colhead{$\chi^2$} & 
\colhead{$\sigma_{\rm res}$} \\ 
\colhead{(keV)} & 
\colhead{(MG)} & 
\colhead{($^{\circ}$)} & 
\colhead{ } & 
\colhead{ } & 
\colhead{($a_{\rm wd})$\tablenotemark{a}} & 
\colhead{ } & 
\colhead{(mJy)}
}
\startdata
12 & 13.6  & 75 & 2.845 & $-25.178$ & 0.0145 & 0.202 & 0.0432 \\
10 & 13.4  & 75 & 3.267 & $-25.106$ & 0.0172 & 0.224 & 0.0475 \\
 8 & 13.35 & 75 & 3.653 & $-25.044$ & 0.0198 & 0.424 & 0.0535 \\
 6 & 13.2  & 75 & 4.342 & $-24.919$ & 0.0264 & 0.728 & 0.0737 
\enddata
\tablenotetext{a}{The cyclotron emitting area $a_{\rm cyc}$ determined 
by the scale factor $A$ is given in units of the projected WD surface 
area $a_{\rm wd}=\pi R_{\rm wd}^2$.}
\end{deluxetable}

\begin{deluxetable}{lll}
\tablewidth{0pt}
\tabletypesize{\footnotesize}
\tablecaption{Shared model parameters \label{t:shared-params}} 
\tablehead{
\colhead{Component} & 
\colhead{Parameter} & 
\colhead{Value}
}
\startdata
System: & Orbital Period, $P_{\rm orb}$ (min) & 81.022932(8) [5] \\
        & Inclination, $i$ ($^{\circ}$) & 55(5) [5] \\
        & Distance, $d$ (pc) & 132 [6] \\
WD:     & Temperature, $T_{\rm wd}$ (1000 K) & 9.5(0.5) [1] \\
        & Mass, $M_{\rm wd}$ ($M_{\odot}$) & 0.6 [1] \\
        & Radius, $R_{\rm wd}$ ($R_{\odot}$) & 0.0125 [2] \\
SS:     & Template Star & 2MASS J15074769$-$1627386 \\
        & Spectral Type & L5 [2,3] \\
        & Temperature, $T_{2}$ (1000 K) & 1.7 [2] \\
        & Mass, $M_2$ ($M_{\odot}$) & 0.055 [4] \\
        & Radius, $R_{2}$ ($R_{\odot}$) & 0.1 [2] \\
CBD:    & Optical Depth Prescription & Thin \\
        & Temperature Profile Exponent & 0.75 \\
        & Disk Annulus Treatment & Equal mass \\
        & Constant Height, $h_{\rm cbd}$ ($R_{\rm wd}$) & 0.00167 \\
        & Grain Density, $\rho_{\rm grain}$ (g cm$^{-3}$) & 3.0 \\
        & Grain Radius, $r_{\rm grain}$ ($\mu$m) & 1 
\enddata
\tablerefs{
[1] = \citet{beuermann00},
[2] = \citet{brinkworth07} (and references therein),
[3] = \citet{HC01},
[4] = \citet{howell06b},
[5] = \citet{piirola87}, and
[6] = \citet{thor03}.
}
\end{deluxetable}

\begin{deluxetable}{llll}
\tablewidth{0pt}
\tabletypesize{\footnotesize}
\tablecaption{Model-specific parameters \label{t:specific-params}} 
\tablehead{
\colhead{Component} & 
\colhead{Parameter} & 
\multicolumn{2}{c}{Values:} \\
\colhead{ } & 
\colhead{ } & 
\colhead{Model 1} &
\colhead{Model 2} 
}
\startdata
CYC:    & WD Magnetic Field, $B$ (MG) & 13.4 & 13.4 \\
        & Electron Temperature, $kT$ (keV) & 10 & 10 \\
        & Angle to Magnetic Field, $\theta$ ($^{\circ}$) & 75 & 75 \\
        & Size Parameter, $\log(\Lambda)$ & 3.267 & 3.301 \\
        & Scale Factor, $\log(A)$ & $-25.106$ & $-25.220$ \\
        & Emitting Area, $a_{\rm cyc}$ ($a_{\rm wd}$) & 0.0172 & 0.0132 \\
        & Model Goodness, $\chi^2$ & 0.224 & 0.286 \\
        & Model Residuals, $\sigma_{\rm res}$ (mJy) & 0.0475 & 0.0522 \\
CBD:    & Inner Edge Temperature, $T_{\rm in}$ (K) & 830 & 450 \\
        & Inner Edge Radius, $R_{\rm in}$ ($R_{\rm wd}$) & 73  & 73 \\
        & Outer Edge Temperature, $T_{\rm in}$ (K) & 20  & 20 \\
        & Outer Edge Radius, $R_{\rm in}$ ($R_{\rm wd}$) & 10482  & 4634 \\
        & Total Mass, $M_{\rm cbd}$ ($10^{20}$ g) & 9.55 & 17.3 \\
BB:     & Temperature, $T$ (K) & \nodata & 1000 \\
        & Radius of emitting area, $R_{\rm bb}$ ($R_{\rm wd}$) & \nodata & 45.6
\enddata
\end{deluxetable}



\clearpage

\end{document}